\documentclass[12pt]{article}

\makeatletter
\def\underbracket{%
    \@ifnextchar[{\@underbracket}{\@underbracket [\@bracketheight]}%
}
\def\@underbracket[#1]{%
    \@ifnextchar[{\@under@bracket[#1]}{\@under@bracket[#1][0.4em]}%
}
\def\@under@bracket[#1][#2]#3{
    \mathop{\vtop{\m@th \ialign {##\crcr $\hfil \displaystyle {#3}\hfil $
    \crcr \noalign {\kern 3\p@ \nointerlineskip }\upbracketfill {#1}{#2}
     \crcr \noalign {\kern 3\p@ }}}}\limits}
\def\upbracketfill#1#2{$\m@th \setbox \z@ \hbox {$\braceld$}
                   \edef \@bracketheight{\the \ht \z@}\bracketend{#1}{#2}
                   \leaders \vrule \@height #1 \@depth \z@ \hfill
                   \leaders \vrule \@height #1 \@depth \z@ \hfill
                   \bracketend{#1}{#2}$}
\def\bracketend#1#2{\vrule height #2 width #1\relax}
\makeatother

\usepackage{graphicx,epsfig}
\def\lsim{\mathrel{\raise.3ex\hbox{$<$\kern-.75em\lower1ex\hbox{$\sim$}}}}
\def\gsim{\mathrel{\raise.3ex\hbox{$>$\kern-.75em\lower1ex\hbox{$\sim$}}}}
\newcommand{ \slashchar }[1]{\setbox0=\hbox{$#1$}   % set a box for #1
   \dimen0=\wd0                                     % and get its size
   \setbox1=\hbox{/} \dimen1=\wd1                   % get size of /
   \ifdim\dimen0>\dimen1                            % #1 is bigger
      \rlap{\hbox to \dimen0{\hfil/\hfil}}          % so center / in box
      #1                                            % and print #1
   \else                                            % / is bigger
      \rlap{\hbox to \dimen1{\hfil$#1$\hfil}}       % so center #1
      /                                             % and print /
   \fi}                                             %
\def\be{\begin{equation}}
\def\ee{\end{equation}}
\def\bea{\begin{eqnarray}}
\def\eea{\end{eqnarray}}
\def\bec{\begin{center}}
\def\eec{\end{center}}
\def\atversim#1#2{\lower0.7ex\vbox{\baselineskip\zatskip\lineskip\zatskip
  \lineskiplimit 0pt\ialign{$\matth#1\hfil##\hfil$\crcr#2\crcr\sim\crcr}}}

\renewcommand{\thefootnote}{\fnsymbol{footnote}}

\newcounter{appendixc}
\newcounter{subappendixc}[appendixc]
\newcounter{subsubappendixc}[subappendixc]

\renewcommand{\appendix}[1] {\vspace*{0.6cm}
        \refstepcounter{appendixc}
        \setcounter{figure}{0}
        \setcounter{table}{0}
        \setcounter{equation}{0}
        \renewcommand{\thefigure}{\Alph{appendixc}.\arabic{figure}}
        \renewcommand{\thetable}{\Alph{appendixc}.\arabic{table}}
        \renewcommand{\theappendixc}{\Alph{appendixc}}
        \renewcommand{\theequation}{\Alph{appendixc}.\arabic{equation}}
        \noindent{\bf Appendix \theappendixc #1}\par\vspace*{0.4cm}}

\begin{document}
\begin{titlepage}
%\rightline{\vbox{\halign{&#\hfil\cr &GUCAS-SPS-05-06 \cr
%&hep-ph/0512214\cr }}} \vskip .5in
\begin{center}

{\Large\bf $0^{-+}$ Trigluon Glueball and its Implication for a
Recent BES Observation}
\vskip .5cm

\normalsize {\bf  Gang Hao\footnote{Email:
hao\_gang@mails.gucas.ac.cn}}$^{1}$, {\bf Cong-Feng
Qiao\footnote{Email: qiaocf@gucas.ac.cn}}$^{2,1}$,
{\bf Ailin Zhang\footnote{Email: zhangal@staff.shu.edu.cn}}$^{3}$\\
\vskip .3cm

$^1$ Dept. of Physics, Graduate School of the Chinese
Academy of Sciences,\\
YuQuan Road 19A, Beijing 100049, China \\
%\vskip .3cm

$^2$ CCAST(World Lab.), P.O. Box 8730, Beijing 100080, China\\
%\vskip .3cm
 $^3$ Department of Physics, Shanghai University,\\
Shangda Road 99, Shanghai 200444, China \vskip 1.5cm

\end{center}

\begin{abstract}
\normalsize

We calculate the mass of $0^{-+}$ triple-valence-gluon resonance,
the trigluon glueball, with QCD sum rules. Its mass is found to be
approximately in the region between $1.9$ GeV and $2.7$ GeV with
some theoretical uncertainties. Moreover, it is likely that the new
BES measurement of the $p \bar{p}$ enhancement near threshold in the
$J/\psi$ decays exhibits the behavior of this trigluon state. Our
analyzes favor the baryonium-gluonium mixing picture for the BES
observation.

\end{abstract}
\vspace{0.2cm}
\renewcommand{\thefootnote}{\arabic{footnote}}
\end{titlepage}

%%%%%%%%%%%%%%%%%%%%%%%%%%%%%%%%%%%%%%%%%%%%%%%%%%%%%%%%%%%%%%%%%%%%%%%%
So far, one of the mysteries remaining in the Standard Model is the
undiscovered elusive gluonium, or glueball, which is theoretically
expected based on the non-Abelian and confinement natures of Quantum
Chromodynamics(QCD). Although glueball has been searched for by
experiments for many years, there has been no clear evidence yet.
Due to gluon self-interaction, the valence gluons inside the
gluonium, suppose it exist, should be heavily "dressed" through the
vacuum fluctuation. This kind of non-perturbative interaction,
together with the bound state confinement effects, makes the
theoretical investigation pretty difficult. A even worse situation
is that there might be a strong glueball-meson mixing, which may
obscure the predicted glueball detection in experiment.

SVZ QCD sum rule \cite{SVZ} has been applied successfully to many
hadron phenomena, such as hadron spectrum and hadron decays. In this
method, different interpolating currents are constructed
corresponding to different hadrons, then the sum rule is established
by constructing a correlation function and matching its operator
product expansion (OPE) to its hadronic saturation. In the
correlation function, the short- and long-distance strong
interactions are separated by means of the OPE, where the former is
perturbative calculated, whereas the latter are treated as
non-perturbative fundamental parameters. Although this approach is
plagued by its large theoretical uncertainties, it gives a model
independent treatment of hadrons. At least, its results are
qualitatively reliable and meaningful. Especially, before the
lattice simulation, which starts from the QCD first principle, can
make a more precise prediction on the physics of hadronic states,
e.g. glueball, these kinds of studies are necessary.

The two-valence-gluon resonances, the two-gluon glueballs(or
bigluonium), have been studied extensively in the literatures. With
QCD sum rules, the mass of $0^{++}$ scalar glueball
\cite{gball1,gball2,
gball3,gball4,gball5,gball6,gball7,gball8,gball9,gball10,gball11}
and $2^{++}$ tensor glueball were calculated \cite{gball12,gball13},
and their masses were predicted to be $1.6(\pm 0.3)$ GeV and
$1.7(\pm 0.5)$ GeV, respectively. According to lattice
simulations\cite{lattice1,lattice2,lattice3,lattice4,lattice5}, the
$0^{-+}$ glueball should be the third lightest pure gluonic state.
Relevant theoretical investigations on this $0^{-+}$ state have been
performed, but the studies are limited only to the two-valence-gluon
glueball (the leading Fock state in Fock space
expansion)\cite{gball8,gball12,gball14}. It should be noted that the
$0^{-+}$ glueball can also be constructed by tree-valence gluons,
which is different from the next-to-leading Fock state of the
bigluonium in Fock space expansion. About this point, one can easily
find similar cases in the quark hadron sector.

In this work we will calculate the three-gluon $0^{-+}$ glueball
mass by means of the QCD sum rules. The only independent triple
gluon Lorentz irreducible and gauge invariant interpolating current
for $0^{-+}$ glueball is found to be
\begin{eqnarray}
j(x)=g_s^3f^{abc}\widetilde{G}^{a}_{\mu\nu}(x)\widetilde{G}^{b}_{\nu\rho}(x)
\widetilde{G}^{c}_{\rho\mu}(x)\; ,
\end{eqnarray}
where,
\begin{eqnarray}
\qquad\widetilde{G}^{a}_{\mu\nu}(x)=\frac{1}{2}\epsilon_{\mu\nu\kappa\tau}
{G}^{a}_{\kappa\tau}(x)\; ,
\end{eqnarray}
$f^{abc}$ is the antisymmetric SU(3) structure constant,
$G_{\mu\nu}^{a}$ is the gluon field strength tensor, and three
$\widetilde{G}$s are introduced here for the convenience of
calculation.
\begin{figure}[htdp]
\begin{center}
\psfig{figure=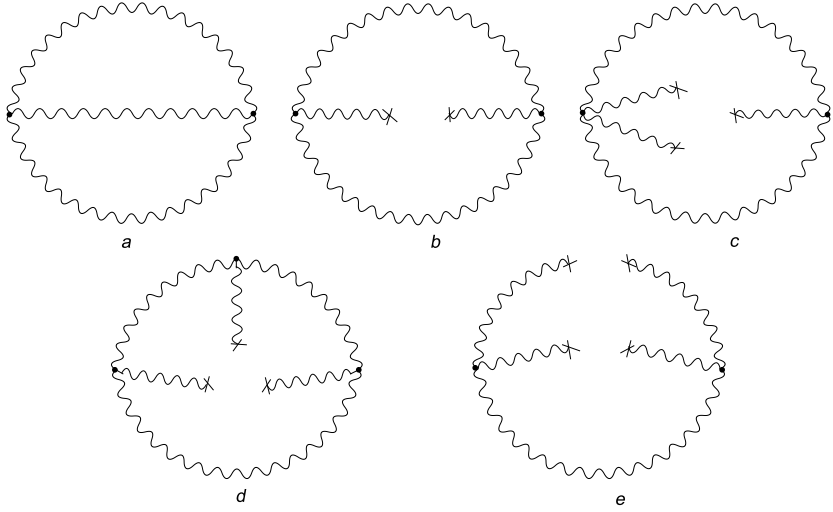,width=11cm,height=6.8cm %angle=-90
}
%\vspace{-18.5cm}
\caption{The typical Feynman diagrams of three-gluon glueball in the
scheme of QCD sum rules. (a) diagram for the perturbative term; (b)
for the two gluon condensate terms; (c) and (d) for the three-gluon
condensate terms; (e) for the four-gluon condensate terms.}
\label{fig1}
\end{center}
\end{figure}
\par
The two-point correlation function for the sum rules is then
constructed as:
\begin{equation}
\Pi(q^2)=i\int d^{4}x\,e^{iq\cdot
x}\langle{\bf{0}}|T\{j(x)j(0)\}|{\bf{0}}\rangle \; ,
\end{equation}
where $|{\bf{0}}\rangle$ represents the physical vacuum. The Feynman
diagrams of the correlation function in our calculation are shown in
Figure 1. After a lengthy calculation, some formula are given in the
appendix, the final result of the OPE for the correlation function
up to order of four-gluon condensate is obtained
\begin{equation}\label{ope1}
\Pi(Q^2)=a_0Q^8\ln \frac{Q^2}{\mu^2}+b_0Q^4\langle\alpha_s
G^2\rangle+(c_0+c_1\ln\frac{Q^2}{\mu^2})Q^2\langle g_s
G^3\rangle+d_0^{(a)}\langle\mathcal{O}_{8a}\rangle+d_0^{(b)}
\langle\mathcal{O}_{8b}\rangle \; ,
\end{equation}
where $\mu$ is the renormalization scale, $Q^2 \equiv -q^2>0$, and
the coefficients are
\begin{equation}
a_0=-\frac{6}{5}\frac{\alpha_s^3}{4\pi},b_0=18\pi\alpha_s^2,c_0=-189\pi\alpha_s^3,
c_1=54\pi\alpha_s^3,d_0^{(a)}=0,d_0^{(b)}=-9(4\pi)^3\alpha_s\; .
\end{equation}
The condensates are defined as
\begin{eqnarray}
\langle\alpha_sG^2\rangle=\langle\alpha_sG^a_{\mu\nu}G^a_{\mu\nu}\rangle\; , \\
\langle g_sG^3\rangle=\langle g_sf^{abc}G^a_{\mu\nu}G^b_{\nu\rho}G^c_{\rho\mu}\rangle\; ,\\
\langle\mathcal{O}_{8a}\rangle=\langle(\alpha_sf^{abc}G^a_{\mu\nu}G^b_{\rho\sigma})^2\rangle\; ,\\
\langle\mathcal{O}_{8b}\rangle=\langle(\alpha_sf^{abc}G^a_{\mu\nu}G^b_{\nu\rho})^2\rangle\; .
\end{eqnarray}

With the exact form of Eq. (\ref{ope1}), the dispersion relation
with subtractions reads
\begin{equation}
\Pi(Q^2)=\frac{(Q^2)^4}{\pi}\int_0^{+\infty}
ds\frac{\mathrm{Im}\Pi(s)}{s^4(s+Q^2)}+\Big(\Pi(0)-Q^2\Pi'(0)+\frac{1}{2}Q^4\Pi''(0)
-\frac{1}{6}Q^6\Pi'''(0)\Big)\; ,
\end{equation}
where $\Pi(0),~\Pi'(0)$ and $\Pi''(0)$ are constants relevant to the
correlation function at origin.

On the other hand, in the narrow width approximation, the imaginary
part of the correlation function can be saturated as:
\begin{equation}
\frac{1}{\pi}\mathrm{Im}\Pi(s)=f^2M^8\delta(s-M^2)+\rho(s)\theta(s-s_0)\;
,
\end{equation}
where $\rho(s)$ is the spectral function of excited states and
continuum states above the continuum threshold $s_0$, $M$ is the
mass of the resonance (glueball) and $f$ is the coupling defined by:
\begin{equation}
\langle 0|j(0)|G\rangle = fM^4.
\end{equation}

To take control of the contribution of high order condensates in OPE
and the contribution of high excited resonance and continuum states
in the hadron side, the Borel transformation is performed
\begin{equation}
\hat{B}_{\tau}\equiv \lim_{Q^2\rightarrow \infty,n\rightarrow \infty
\atop{Q^2\over n}=
{1\over\tau}}\frac{(-Q^2)^n}{(n-1)!}\left(\frac{d}{dQ^2}\right)^n
\end{equation}
to the sum rule, and the moment $R_k$ is obtained
\begin{eqnarray}
R_k(\tau,s_0)&=&\frac{1}{\tau}\hat{B}_\tau\bigg[(-Q^2)^k \
\Big(\Pi(Q^2)-\frac{(Q^2)^4}{\pi}\int_{s_0}^\infty ds
\frac{\rho(s)}{s^4(s+Q^2)}\Big)\bigg]
\nonumber\\
&=&\frac{1}{\pi}\int_{0}^{\infty}s^k
e^{-s\tau}\mathrm{Im}\Pi(s)ds-\int_{s_0}^{\infty}s^k
e^{-s\tau}\rho(s)ds \; .
\end{eqnarray}
Employing the quark-hadron duality approximation
\begin{equation}
\int_{s_0}^{\infty}s^k e^{-s\tau}\rho(s)ds \simeq
\frac{1}{\pi}\int_{s_0}^{\infty}s^k
e^{-s\tau}\mathrm{Im\Pi^{pert}}(s)ds \; ,
\end{equation}
we have
\begin{equation}
R_k(\tau,s_0)=\frac{1}{\pi}\int_{0}^{s_0}s^k
e^{-s\tau}\mathrm{Im\Pi}(s)ds \; .
\end{equation}
Then, the glueball mass can be extracted from the ratio of the
above moments, i.e.,
\begin{equation}
M^2(s_0,\tau)=\frac{R_k}{R_{k-1}}\label{ratio}\; .
\end{equation}
The first two moments in Eq.(\ref{ratio}) are
\begin{eqnarray}
R_0(s_0,\tau)&=&-\frac{24a_0}{\tau^5}[1-\rho_4(s_0\tau)]+\frac{c_1}{\tau^2}\langle
g_sG^3\rangle\; ,\\
R_1(s_0,\tau)&=&-\frac{120a_0}{\tau^6}[1-\rho_5(s_0\tau)]+\frac{2c_1}{\tau^3}\langle
g_sG^3\rangle ,
\end{eqnarray}
where
\begin{eqnarray}
\rho_k(x)&=&e^{-x}\sum^k_{n=0}\frac{x^n}{n!} \; .
\end{eqnarray}
From what obtained above, we can easily get the analytic function of
glueball mass relying on the Borel transformation parameter $\tau$
and threshold $s_0$.

For numerical analysis, the input parameters in our calculation are
taken to be
\begin{eqnarray}
\langle\alpha_s G^2\rangle&=&0.06~GeV^4 \; , \nonumber\\
\langle g_s G^3\rangle&=&(0.27 ~GeV^2)\langle\alpha_s
G^2\rangle\nonumber \; , \\
\langle\mathcal{O}_{8b}\rangle&=&\frac{1}{16}\langle\alpha_s
G^2\rangle^2\nonumber \; , \\
\Lambda_{\overline{MS}}&=&300\mathrm{MeV}\nonumber  \; , \\
\alpha_s&=&\frac{-4\pi}{11 \ln(\tau\Lambda_{\overline{MS}}^2)}\; .
\end{eqnarray}

The three gluon condensate in dilute gas instanton model is used
here for lack of direct extraction from experimental data. It is
found that the variation of three gluon condensate changes the
predicted glueball mass small. Other parameters are commonly used in
literatures. The numerical dependence of glueball mass on the $\tau$
is presented in Figure 2. There is not a exact stable window in the
figure. However, the glueball mass decreases slowly with the
increase of Borel parameter $\tau$, an approximate glueball mass
could be determined. The excited states and continuum threshold
parameter $s_0=6.5$ $\rm{GeV}^2$ is chosen according to the method
in Ref. \cite{gball9}. To determine this $s_0$, the suitable region
for $\tau$ lies between $0.2~GeV^{-2}$ and $0.7~ GeV^{-2}$. At every
$\tau$ in this region, the predicted glueball mass varies slowly
with $s_0$. Therefore, the glueball mass could be approximately
regarded as falling in the scope of $1.9$ GeV to $2.7$ GeV.

The mass of $0^{-+}$ three gluon glueball obtained here is not as
large as one expected (the mass of $0^{++}$ three gluon glueball is
predicted around $3.1$ GeV\cite{narison}, where the numerical
analysis is a little different from ours). There are several
possible reasons for it. One possibility is that we have
under-estimated the mass of three gluon glueball for the theoretical
uncertainties. This under-estimation may originate from several
ways.

In the sum rules constructed with gluons current, two gluon
condensate is expected to give a large contribution, which may
change the mass determination. However, the leading order two gluon
condensate does not enter into the sum rule after Borel
transformation both in our calculation and in Ref. \cite{narison}.
If one wishes to keep two gluon condensate in SVZ sum rules and to
see how large they will contribute, it is necessary to calculate the
next leading order contribution of two gluon condensate or to know
the behavior of the correlation function at origin.

As pointed out in Refs.\cite{shuryak,gball11}, in the pseudoscalar
glueball (two gluons) and $\eta^\prime$ channels, the topological
charge screening contributions to the sum rules are found to be
large. This screening effect generated from the interaction between
the topological charges is expected to exist also in the
pseudoscalar three-gluon channel though its exact contribution has
not yet been studied.

In addition, in the pseudoscalar flavor singlet channel, the mixing
among glueballs (three-gluon and two-gluon glueball) and normal
mesons ($\eta$ and $\eta^\prime$) may change our result. All these
contributions are beyond our scope in this paper. Certainly, the
variations of $\tau$, $s_0$ and other input parameters will result
in some uncertainties also.

Another possibility is that we have touched the reality. Our result
for the $0^{-+}$ glueball is consistent with the recent lattice
result ($2.56$ GeV)\cite{lattice5}. As we know, in the lattice
calculation, the predicted $0^{-+}$ glueball includes two gluon
glueball, three gluon glueball and glueball with more than three
gluons. If this is the truth, it implies that one additional
dynamical gluon attributes less than one GeV to the glueball mass,
not as what normally think in the hybrid case. To clarify this point
of view, more investigations in other models are required.
\begin{figure}[htdp]
\begin{center}
\psfig{figure=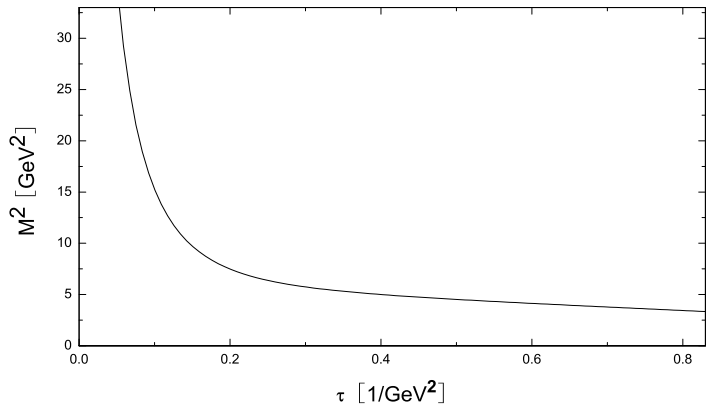,width=11cm,height=7cm %angle=-90
}
%\vspace{-18.5cm}
\caption{The trigluon glueball mass square dependence on the Borel
parameter $\tau$. The excited states and continuum threshold
parameter $s_0$ is taken to be 6.5 $\rm{GeV^2}$ } \label{fig2}
\end{center}
\end{figure}
\par
In recently, the BES Collaboration has observed an enhancement at
the $p \bar{p}$ threshold with the quantum number $0^{-+}$
\cite{BES} and a resonance $X(1835)$\cite{BES2} with $J^{PC}$
undetermined. It is still unclear whether these two enhancements are
the same. These observations have stirred many explanations such as
the $p\bar{p}$ baryonium state \cite{gjding}, or the gluonic state
\cite{bali,xghe, nk}. In our understanding, the $p \bar{p}$
enhancement is probably a mixture of baryonium state with the
glueball(both two- and three-gluon states), which is also proposed
in Ref.\cite{gjding}. However, we think that if the
baryonium-gluonium mixing picture is the reality for the BES
observation\cite{BES}, there should be large component of
three-gluon glueball inside. The arguments for this idea are:

1) The observed $p\bar{p}$ enhancement is close to the mass scope
of $0^{-+}$ glueball in our calculation.

2) The new observation exhibits large $\gamma + p\bar{p}$(and also
three mesons) branching ratio relative to the two meson production
processes in comparison with the decays of other known states.
This unique character may hint that the new state has relatively
strong coupling to six-quark final states \cite{xyshen}. In the
picture of the three gluon glueball, this is easy to be
understood, since the three valence gluons can split into six
light quarks easily and form $p\bar{p}$ final states.

3) Theoretically, the rate of $J/\psi$ radiative decays to three
gluons is in a similar order of magnitude as to two gluons, while
we should enforce a proper physical cut to avoid the infrared
singularities to draw such a conclusion \cite{FDC}.

4) The missing $\Lambda\bar{\Lambda}$ threshold enhancement implies
the obvious flavor SU(3) breaking and favors the idea that what BES
observed is a mixed state of $p\bar{p}$ baryonium and glueball, and
the latter with a mass closing to the $p\bar{p}$ threshold. The pure
baryonium interpretation meets the difficulty of large SU(3)
breaking, while pure gluonic state interpretation has an overshoot
mass. The three-gluon glueball-baryonium mixing(oscillation) picture
is a simple and natural one to this understanding.

In conclusion, in this paper we calculate the $0^{-+}$ three-gluon
glueball mass by means of QCD sum rules. We find that its mass lies
in the region of $1.9 \sim 2.7$ GeV, which is consistent with the
lattice simulation result. Since our calculation is performed in the
leading order, the instanton and topological charge screening
contributions have not been included in, the mixing effect among
glueballs and normal mesons has not considered, and the input
parameters remain some uncertainties, the mass prediction can not be
very precise. In our leading order analysis, only the
non-perturbatively triple-gluon condensate gives contributions to
the glueball mass, whereas the two- and four-gluon condensates
vanish after the Borel transformation. In further effort on the aim
of making an accurate prediction, one should include the higher
order corrections from both perturbative and non-perturbative
sectors, in which the two- and four-gluon condensates contribute.
The instanton and the topological charge screening contributions
should be taken into account. Furthermore, the mixing effect among
glueballs (two-gluon and three gluon glueball) and normal mesons
($\eta$ and $\eta^\prime$) should be considered.

In fact, what we calculated in this work is partially motivated by
the new observation at BES. From the naive arguments listed above,
we think the new BES observation of an enhancement at $p\bar{p}$
threshold might be a mixed state of baryonium and three-gluon
glueball. Nevertheless, if our conjecture is true, how much the
gluonic, or baryonium, content engages in the observed enhancement
still needs more investigations, in both experiment and theory.

%%%%%%%%%%%%%%%%%%%%%%%%%%%%%%%%%%%%%%%%%%%%%%%%%%%%%%%%%%%%%%%%%%%%%%%%%%%%%%%%
%\vspace{1.0cm}
\par
{\bf Acknowledgments}
C.-F. Qiao acknowledges for the helpful
discussion with colleagues of BES new phenomenon working group and
J. X. Wang. Ailin Zhang acknowledges Ying Chen for the discussion
about lattice results on glueball. The authors thank Dr. J. Latorre
very much for his mention of reference\cite{narison}. This work was
supported in part by the National Natural Science Foundation of
China.
%%%%%%%%%%%%%%%%%%%%%%%%%%%%%%%%%%%%%%%%%%%%%%%%%%%%%%%%%%%%%%%%%%%%%%%%%%%%%%%%

\appendix

\begin{eqnarray*}
\Pi^{pert}(Q^2)&=&6i g_s^6 f^{abc}f^{ijk}\int\! d^4 x\,e^{iq\cdot
x}[\underbracket[1pt][0.4em]{\widetilde{G}^a_{0\mu\nu}(x)\widetilde{G}^{i}_{0\alpha\beta}}\!\!(0)][\underbracket[1pt][0.4em]{\widetilde{G}^b_{0\nu\rho}(x)\widetilde{G}^j_{0\beta\gamma}}\!\!(0)][\underbracket[1pt][0.4em]{\widetilde{G}^c_{0\rho\mu}(x)\widetilde{G}^k_{0\gamma\alpha}}\!\!(0)]\\
&=&-6 g_s^6 f^{abc}f^{abc}\int\!\!\!\int\!\!\!\int\!\frac{d^D
p_1}{(2\pi)^D}\frac{d^D p_2}{(2\pi)^D}\frac{d^D
p_3}{(2\pi)^D}\frac{(2\pi)^D\delta^D(q-p_1-p_2-p_3)}{p_1^2
p_{2}^{2}p_{3}^{2}}\\
& &\times\Big\{(-D^2+8D-14)\left[p_3^2(p_1\cdot
p_2)^2+p_1^2(p_2\cdot p_3)^2+p_2^2(p_1\cdot
p_3)^2\right]\\
& &\ \ \ \ \ +(-D^3+9D^2-29D+32)p_1^2p_2^2p_3^2+(16-6D)(p_1\cdot
p_2)(p_2\cdot
p_3)(p_1\cdot p_3)\Big\}\\
&=&\frac{3\alpha_s^3}{10\pi}
Q^8\left(\frac{1}{2\epsilon}-\gamma_E+\frac{87}{40}-\ln\frac{Q^2}{4\pi\nu^2
}\right)
\end{eqnarray*}
\begin{eqnarray*} \Pi_4(Q^2) & = & 18i g_s^6 f^{abc}f^{ijk}\int\!
d^{D}x\ e^{iq\cdot
x}[\underbracket[1pt][0.4em]{\widetilde{G}^a_{0\mu\nu}(x)\widetilde{G}^{i}_{0\alpha\beta}}\!\!(0)][\underbracket[1pt][0.4em]{\widetilde{G}^b_{0\nu\rho}(x)\widetilde{G}^j_{0\beta\gamma}}\!\!(0)]\langle
0 |\widetilde{G}^{c}_{0\rho\mu}(x) \widetilde{G}^{k}_{0\gamma\alpha}(0)|0\rangle \\
& = & 27i g_s^6 \frac{(D-2)}{D(D-1)}\langle
GG\rangle\\
&
&\quad\times\int\!\frac{d^{D}p_{1}}{(2\pi)^D}\frac{1}{{p_1}^{2}{p_2}^{2}}[(D^2-6D+8)(p_1\cdot
               p_2)^{2}+(D^3-7D^2+17D-14){p_1}^{2}{p_2}^{2}]\Big|_{p_2=q-p_1}\\
               &=&18\pi\alpha^2_s Q^4 \langle \alpha_s G^2
               \rangle\\
\end{eqnarray*}
\begin{equation}
\Pi_6(Q^2)=\sum_{n=1}^{13} S_n(Q^2)
\end{equation}
\begin{eqnarray*}
S_1(Q^2)&=&18ig_s^7 f^{abc} f^{ijk} f^{amn} \\
& &\quad\times\int\!d^4 x \,e^{iq\cdot
x}[\underbracket[1pt][0.4em]{\widetilde{G}^b_{0\nu\rho}(x)
\widetilde{G}^j_{0\beta\gamma}}\!\!(0)][\underbracket[1pt][0.4em]{\widetilde{G}^c_{0\rho\mu}(x)
\widetilde{G}^k_{0\gamma\alpha}}\!\!(0)]\frac{1}{4}\epsilon_{\mu\nu\kappa\tau}\epsilon_{\alpha\beta\eta\xi}\langle0|A^m_\kappa(x)
A^n_\tau(x)
G^i_{0\eta\xi}(0)|0\rangle\\
&=&i\frac{27}{2}\frac{g_s^6\langle g_s GGG
\rangle}{D(D-1)(D-2)}\int\!\frac{d^D
p_1}{(2\pi)^D}\frac{1}{p_1^4 p^2_2}\\
& &\times\left[(D^4-12D^3+56D^2-116D+88)p_1^2
p_2^2 +(D^4-10D^3+24D^2-32)(p_1\cdot p_2)^2\right]\Big|_{p_2=q-p_1}\\
&=&54\pi \alpha_s^3 Q^2\left(\frac{1}{\epsilon}-\gamma_E-\ln\frac{Q^2}{4\pi\nu^2}+\frac{3}{2}\right)\langle g_s GGG \rangle\\
\end{eqnarray*}
\begin{eqnarray*}
S_2(Q^2)&=&18ig_s^7 f^{abc} f^{ijk} f^{amn}\\
& &\quad\times \int\!d^4 x \,e^{iq\cdot
x}[\frac{1}{2}\epsilon_{\mu\nu\kappa\tau}\underbracket[1pt][0.4em]{A^m_\kappa(x)
\widetilde{G}^j_{0\beta\gamma}}\!\!(0)][\frac{1}{2}\epsilon_{\alpha\beta\eta\xi}\underbracket[1pt][0.4em]{A^n_\tau(x)
\widetilde{G}^k_{0\gamma\alpha}}\!\!(0)]\langle0|\widetilde{G}^b_{0\nu\rho}(x)\widetilde{G}^c_{0\rho\mu}(x)
G^i_{0\eta\xi}(0)|0\rangle\\
&=&\frac{-i54g_s^6\langle g_s GGG
\rangle}{D(D-1)(D-2)}\int\frac{d^D
p_1}{(2\pi)^D}\frac{(D-1)(D-2)^2 p_1\cdot(q-p_1)}{p_1^2 (q-p_1)^2}\\
&=&-54\pi\alpha_s^3 Q^2\left(\frac{1}{\epsilon}-\gamma_E-\ln\frac{Q^2}{4\pi\nu^2}+\frac{3}{2}\right)\langle g_s GGG \rangle\\
\end{eqnarray*}
\begin{eqnarray*}
S_3(Q^2)&=&72ig_s^7 f^{abc} f^{ijk} f^{amn} \\
& &\quad\times\int\!d^4 x \,e^{iq\cdot
x}[\frac{1}{2}\epsilon_{\mu\nu\kappa\tau}\underbracket[1pt][0.4em]{A^m_\kappa(x)\widetilde{
G}^j_{0\beta\gamma}}\!\!(0)][\underbracket[1pt][0.4em]{\widetilde{G}^c_{0\rho\mu}(x)
\widetilde{G}^k_{0\gamma\alpha}}\!\!(0)]\langle0|A^n_\tau(x)\widetilde{G}^b_{0\nu\rho}(x)
\widetilde{G}^i_{0\alpha\beta}(0)|0\rangle\\
&=&\frac{i27g_s^6\langle g_s GGG \rangle
}{D(D-1)(D-2)}\int\frac{d^D
p_1}{(2\pi)^D}\frac{(D^3-7D^2+14D-8)(p_1\cdot p_2)}{p_1^2 p_2^2}\bigg|_{p_2=q-p_1}\\
&=&-27\pi\alpha_s^3 Q^2\langle
g_s GGG \rangle\\
\end{eqnarray*}
\begin{eqnarray*}
S_4(Q^2)&=&18ig_s^7 f^{abc} f^{ijk} f^{amn} \\
& &\quad\times\int\!d^4 x \,e^{iq\cdot
x}[\underbracket[1pt][0.4em]{\widetilde{G}^b_{0\nu\rho}(x)
\widetilde{G}^j_{0\beta\gamma}}\!\!(0)][\underbracket[1pt][0.4em]{\widetilde{G}^c_{0\rho\mu}(x)
\widetilde{G}^k_{0\gamma\alpha}}\!\!(0)]\frac{1}{4}\epsilon_{\mu\nu\kappa\tau}\epsilon_{\alpha\beta\eta\xi}\langle0|G^a_{0\kappa\tau}(x)A^m_\eta(0)
A^n_\xi(0)
|0\rangle\\
&=&0\\
\end{eqnarray*}
\begin{eqnarray*}
S_5(Q^2)&=&18ig_s^7 f^{abc} f^{ijk} f^{amn} \\
& &\quad\times\int\!d^4 x \,e^{iq\cdot
x}[\frac{1}{2}\epsilon_{\alpha\beta\eta\xi}\underbracket[1pt][0.4em]{A^m_\eta(0)
\widetilde{G}^b_{0\nu\rho}}\!\!(x)][\frac{1}{2}\epsilon_{\mu\nu\kappa\tau}\underbracket[1pt][0.4em]{A^n_\beta(0)
\widetilde{G}^c_{0\rho\mu}}\!\!(x)]\langle0|G^a_{0\kappa\tau}(x)\widetilde{G}^j_{0\beta\gamma}(0)
\widetilde{G}^k_{0\gamma\alpha}(0)|0\rangle\\
&=&-i54g_s^6\langle g_s GGG \rangle\frac{D-2}{D}\int\frac{d^D
p_1}{(2\pi)^D}\frac{p_1\cdot(q-p_1)}{p_1^2 (q-p_1)^2}\\
&=&-54\pi\alpha_s^3 Q^2\left(\frac{1}{\epsilon}-\gamma_E-\ln\frac{Q^2}{4\pi\nu^2}+\frac{3}{2}\right)\langle g_s GGG \rangle\\
\end{eqnarray*}
\begin{eqnarray*}
S_6(Q^2)&=&72ig_s^7 f^{abc} f^{ijk} f^{amn} \\
& &\quad\times\int\!d^4 x \,e^{iq\cdot
x}[\frac{1}{2}\epsilon_{\alpha\beta\eta\xi}\underbracket[1pt][0.4em]{A^m_\eta(0)
\widetilde{G}^b_{0\nu\rho}}\!\!(x)][\underbracket[1pt][0.4em]{\widetilde{G}^k_{0\gamma\alpha}(0)
\widetilde{G}^c_{0\rho\mu}}\!\!(x)]\langle0|A^n_\xi(0)\widetilde{G}^j_{0\beta\gamma}(0)
\widetilde{G}^a_{0\mu\nu}(x)|0\rangle\\
&=&0\\
\end{eqnarray*}
\begin{eqnarray*}
S_7(Q^2)&=&36g_s^7 f^{abc}f^{ijk}f^{dst}\int\!\!\!\int\!d^4 x d^4
y\, e^{iq\cdot x}\\
& &\quad\quad\quad\times[\underbracket[1pt][0.4em]{\widetilde{G}^a_{0\mu\nu}(x)\widetilde{G}^i_{0\alpha\beta}}\!\!(0)][\underbracket[1pt][0.4em]{\widetilde{G}^b_{0\nu\rho}(x)A^s_{\sigma}}\!\!(y)][\underbracket[1pt][0.4em]{A^t_{\delta}(y)\widetilde{G}^j_{0\beta\gamma}}\!\!(0)]\langle0|\widetilde{G}^c_{0\rho\mu}(x)\widetilde{G}^k_{0\gamma\alpha}(0)G^d_{0\sigma\delta}(y)|0\rangle\\
&=&\frac{-i54g_s^7\langle GGG\rangle}{D(D-1)(D-2)}\int\!\frac{d^D
p_2}{(2\pi)^D}\frac{1}{p_1^2 p_2^4}[2(2D-5)p_1^2 p_2^2
+2(D-4)(p_1\cdot p_2)^2]\bigg|_{p_1=q-p_2}\\
&=&-27\pi\alpha_s^3 Q^2\langle g_s GGG\rangle
\end{eqnarray*}
\begin{eqnarray*}
S_8(Q^2)&=&36g_s^7 f^{abc}f^{ijk}f^{dst}\int\!\!\!\int\!d^4 x d^4
y\, e^{iq\cdot x}\\
& &\quad\quad\quad\times[\underbracket[1pt][0.4em]{\widetilde{G}^a_{0\mu\nu}(x)\widetilde{G}^i_{0\alpha\beta}}\!\!(0)][\underbracket[1pt][0.4em]{\widetilde{G}^b_{0\nu\rho}(x)A^s_{\sigma}}\!\!(y)][\underbracket[1pt][0.4em]{G^d_{0\sigma\delta}(y)\widetilde{G}^j_{0\beta\gamma}}\!\!(0)]\langle0|\widetilde{G}^c_{0\rho\mu}(x)\widetilde{G}^k_{0\gamma\alpha}(0)A^t_{\delta}(y)|0\rangle\\
&=&\frac{-27i g_s^7\langle GGG\rangle}{D(D-1)(D-2)}
\int\!\frac{d^D p_2}{(2\pi)^D}\frac{1}{p_1^2
p_2^6}\left[4(D-4) (p_2\cdot p_1)^2p_2^2+4(2D-5)p_1^2 p_2^4\right]|_{p_1=q-p_2}\\
&=&-27\pi\alpha_s^3 Q^2\langle g_s GGG\rangle
\end{eqnarray*}
\begin{eqnarray*}
S_9(Q^2)&=&36g_s^7 f^{abc}f^{ijk}f^{dst}\int\!\!\!\int\!d^4 x d^4
y\, e^{iq\cdot x}\\
& &\quad\quad\quad\times[\underbracket[1pt][0.4em]{\widetilde{G}^a_{0\mu\nu}(x)\widetilde{G}^i_{0\alpha\beta}}\!\!(0)][\underbracket[1pt][0.4em]{\widetilde{G}^b_{0\nu\rho}(x)G^d_{0\sigma\delta}}\!\!(y)][\underbracket[1pt][0.4em]{A^t_{\delta}(y)\widetilde{G}^j_{0\beta\gamma}}\!\!(0)]\langle0|\widetilde{G}^c_{0\rho\mu}(x)\widetilde{G}^k_{0\gamma\alpha}(0)A^s_{\sigma}(y)|0\rangle\\
&=&\frac{27ig_s^7\langle GGG\rangle}{D(D-1)(D-2)} \int\!\frac{d^D
p_2}{(2\pi)^D}\frac{1}{p_1^2
p_2^6}\left[4(4-D) (p_2\cdot p_1)^2p_2^2+4(5-2D)p_1^2 p_2^4\right]\bigg|_{p_1=q-p_2}\\
&=&-27\pi\alpha_s^3 Q^2\langle g_s GGG\rangle
\end{eqnarray*}
\begin{eqnarray*}
S_{10}(Q^2)&=&18g_s^7 f^{abc}f^{ijk}f^{dst}\int\!\!\!\int\!d^4 x
d^4
y\, e^{iq\cdot x}\\
& &\quad\quad\quad\times[\underbracket[1pt][0.4em]{\widetilde{G}^i_{0\alpha\beta}(0)A^t_\delta}\!\!(y)][\underbracket[1pt][0.4em]{\widetilde{G}^b_{0\nu\rho}(x)\widetilde{G}^j_{0\beta\gamma}}\!\!(0)][\underbracket[1pt][0.4em]{\widetilde{G}^c_{0\rho\mu}(x)\widetilde{G}^k_{0\gamma\alpha}}\!\!(0)]\langle0|\widetilde{G}^a_{0\mu\nu}(x)G^d_{0\sigma\delta}(y)A^s_{\sigma}(y)|0\rangle\\
&=&0\\
\end{eqnarray*}
\begin{eqnarray*}
S_{11}(Q^2)&=&9g_s^7 f^{abc}f^{ijk}f^{dst}\int\!\!\!\int\!d^4 x
d^4
y\, e^{iq\cdot x}\\
& &\quad\quad\quad\times[\underbracket[1pt][0.4em]{\widetilde{G}^i_{0\alpha\beta}(0)G^d_{0\sigma\delta}}\!\!(y)][\underbracket[1pt][0.4em]{\widetilde{G}^b_{0\nu\rho}(x)\widetilde{G}^j_{0\beta\gamma}}\!\!(0)][\underbracket[1pt][0.4em]{\widetilde{G}^c_{0\rho\mu}(x)\widetilde{G}^k_{0\gamma\alpha}}\!\!(0)]\langle0|\widetilde{G}^a_{0\mu\nu}(x)A^s_{\sigma}(y)A^t_\delta(y)|0\rangle\\
&=&0\\
\end{eqnarray*}
\begin{eqnarray*}
S_{12}(Q^2)&=&18g_s^7 f^{abc}f^{ijk}f^{dst}\int\!\!\!\int\!d^4 x
d^4
y\, e^{iq\cdot x}\\
& &\quad\quad\quad\times[\underbracket[1pt][0.4em]{\widetilde{G}^a_{0\mu\nu}(x)A^t_\delta}\!\!(y)][\underbracket[1pt][0.4em]{\widetilde{G}^b_{0\nu\rho}(x)\widetilde{G}^j_{0\beta\gamma}}\!\!(0)][\underbracket[1pt][0.4em]{\widetilde{G}^c_{0\rho\mu}(x)\widetilde{G}^k_{0\gamma\alpha}}\!\!(0)]\langle0|\widetilde{G}^i_{0\alpha\beta}(0)G^d_{0\sigma\delta}(y)A^s_{\sigma}(y)|0\rangle\\
&=&0\\
\end{eqnarray*}
\begin{eqnarray*}
S_{13}(Q^2)&=&9g_s^7 f^{abc}f^{ijk}f^{dst}\int\!\!\!\int\!d^4 x
d^4
y\, e^{iq\cdot x}\\
& &\quad\quad\quad\times[\underbracket[1pt][0.4em]{\widetilde{G}^a_{0\mu\nu}(x)G^d_{0\sigma\delta}}\!\!(y)][\underbracket[1pt][0.4em]{\widetilde{G}^b_{0\nu\rho}(x)\widetilde{G}^j_{0\beta\gamma}}\!\!(0)][\underbracket[1pt][0.4em]{\widetilde{G}^c_{0\rho\mu}(x)\widetilde{G}^k_{0\gamma\alpha}}\!\!(0)]\langle0|\widetilde{G}^i_{0\alpha\beta}(0)A^s_{\sigma}(y)A^t_\delta(y)|0\rangle\\
&=&0\\
\end{eqnarray*}
\begin{eqnarray*}
\Pi_6(Q^2)&=&-54\pi \alpha_s^3
Q^2\left(\frac{1}{\epsilon}-\gamma_E-\ln\frac{Q^2}{4\pi\nu^2}+\frac{7}{2}\right)\langle
g_s GGG \rangle\\
\end{eqnarray*}
\begin{eqnarray*}
\Pi_8(Q^2)&=&9ig_s^6f^{abc}f^{ijk}\int\! d^{4}x\ e^{iq\cdot
x}[\underbracket[1pt][0.4em]{\widetilde{G}^{a}_{0\mu\nu}(x)\widetilde{G}^{i}_{0\alpha\beta}}\!\!(0)]\langle
0 |\widetilde{G}^{b}_{0\nu\rho}(x)\widetilde{G}^{c}_{0\rho\mu}(x) \widetilde{G}^{j}_{0\beta\gamma}(0)\widetilde{G}^{k}_{0\gamma\alpha}(0)|0\rangle \\
&=&-9(4\pi)^3\alpha_s^3f^{abc}f^{ajk}\langle
G^b_{\mu\nu}G^{c\nu}_{\ \
\sigma}G^{j\mu\rho}G^{k\sigma}_\rho\rangle \\
&=&-9(4\pi)^3\alpha_s\langle (\alpha_s
f^{abc}G^a_{\mu\nu}G^b_{\nu\rho})^2\rangle\\
\end{eqnarray*}
Field strength tensor:
\begin{displaymath}
G^{a}_{\mu\nu}(x)=G^{a}_{0\mu\nu}(x)+g_{s}f^{abc}A^{b}_{\mu}(x)A^{c}_{\nu}(x)
\end{displaymath}
where
\begin{displaymath}
G^{a}_{0\mu\nu}(x)=\partial_{\mu}A^{a}_{\nu}(x)-\partial_{\nu}A^{a}_{\mu}(x)
\end{displaymath}
In coordinate gauge:
\begin{equation}
A^a_\mu(x)\simeq \frac{1}{2}x^\nu G^a_{\nu\mu}(0);\qquad
A^a_\mu(0)\simeq 0
\end{equation}
\begin{displaymath}
G^a_{\mu\nu}(x)=G^a_{0\mu\nu}(0)+\frac{1}{4}g_s f^{abc} x^\rho
x^\sigma G^b_{\rho\mu}(0)G^c_{\sigma\nu}(0)
\end{displaymath}
\begin{equation}
G^a_{\mu\nu}(0)=G^a_{0\mu\nu}(0)=G^a_{0\mu\nu}(x)
\end{equation}
Some contractions:\\
\begin{displaymath}
\underbracket[1pt][0.4em]{G^a_{0\mu\nu}(x)G^i_{0\alpha\beta}}\!\!(y)=\int\!\frac{d^4
p}{(2\pi)^4}\frac{-i\delta^{ai}}{p^2}\Gamma_{\mu\nu\alpha\beta}(p)e^{-ip\cdot(x-y)}
\end{displaymath}
\qquad\qquad\qquad where
\begin{displaymath}
\Gamma_{\mu\nu\alpha\beta}(p)=p_\mu p_\alpha g_{\nu\beta}+p_\nu
p_\beta g_{\mu\alpha}-p_\mu p_\beta g_{\nu\alpha}-p_\nu p_\alpha
g_{\mu\beta}\\
\end{displaymath}
\begin{displaymath}
\underbracket[1pt][0.4em]{A^m_{\mu}(x)G^j_{0\beta\gamma}}\!\!(y)=\int\!\frac{d^4
p}{(2\pi)^4}\frac{\delta^{mj}}{p_1^2}(p_{\beta}g_{\mu\gamma}-p_{\gamma}g_{\mu\beta})e^{-ip\cdot(x-y)}
\end{displaymath}
\begin{displaymath}
\underbracket[1pt][0.4em]{G^j_{0\beta\gamma}(x)A^m_{\mu}}\!\!(y)=\int\!\frac{d^4
p}{(2\pi)^4}\frac{-\delta^{jm}}{p^2}(p_{\beta}g_{\mu\gamma}-p_{\gamma}g_{\mu\beta})e^{-ip\cdot(x-y)}
\end{displaymath}
\begin{displaymath}
\underbracket[1pt][0.4em]{\widetilde{G}^a_{0\mu\nu}(x)\widetilde{G}^i_{0\rho\sigma}}\!\!(y)=\int\!\frac{d^4
p}{(2\pi)^4}\frac{-i\delta^{ai}}{p^2}\widetilde{\Gamma}_{\mu\nu\rho\sigma}(p)e^{-ip\cdot(x-y)}
\end{displaymath}
\qquad\qquad\qquad where
\begin{displaymath}
\widetilde{\Gamma}_{\mu\nu\rho\sigma}(p)=p_\mu p_\rho
g_{\nu\sigma}+p_\nu p_\sigma g_{\mu\rho}-p_\mu p_\sigma
g_{\nu\rho}-p_\nu p_\rho
g_{\mu\sigma}+p^2(g_{\mu\sigma}g_{\nu\rho}-g_{\mu\rho}g_{\nu\sigma})
\end{displaymath}
Some gluon condensates:\\
\begin{displaymath}
\delta^{ab}\langle0|G^a_{\mu\nu}(0)G^b_{\rho\sigma}(0)|0\rangle=\frac{1}{D(D-1)}(g_{\mu\rho}g_{\nu\sigma}-g_{\mu\sigma}g_{\nu\rho})\langle
GG\rangle
\end{displaymath}
\begin{displaymath}
\delta^{ab}\langle0|\widetilde{G}^a_{\mu\nu}(0)\widetilde{G}^b_{\rho\sigma}(0)|0\rangle=\frac{2-D}{2D(D-1)}(g_{\mu\rho}g_{\nu\sigma}-g_{\mu\sigma}g_{\nu\rho})\langle
GG\rangle
\end{displaymath}
\begin{displaymath}
f^{abc}\langle0|G^a_{\mu\nu}(0)G^b_{\rho\sigma}(0)G^c_{\alpha\beta}(0)|0\rangle=\frac{1}{D(D-1)(D-2)}T^3_{\mu\nu\rho\sigma\alpha\beta}\langle
GGG\rangle
\end{displaymath}
where
\begin{eqnarray*}
T^3_{\mu\nu\rho\sigma\alpha\beta}&=&g_{\mu\rho}g_{\nu\alpha}g_{\sigma\beta}-g_{\mu\rho}g_{\nu\beta}g_{\sigma\alpha}-g_{\mu\sigma}g_{\nu\alpha}g_{\rho\beta}+g_{\mu\sigma}g_{\nu\beta}g_{\rho\alpha}\\
&
&-g_{\mu\alpha}g_{\nu\rho}g_{\sigma\beta}+g_{\mu\alpha}g_{\nu\sigma}g_{\rho\beta}+g_{\mu\beta}g_{\nu\rho}g_{\sigma\alpha}-g_{\mu\beta}g_{\nu\sigma}g_{\rho\alpha}\\
\end{eqnarray*}
Four-gluon condensate:
\begin{eqnarray*}
&
&f^{abe}f^{cde}\langle0|G^a_{\mu\nu}(0)G^b_{\rho\sigma}(0)G^c_{\alpha\beta}(0)G^d_{\gamma\delta}(0)|0\rangle\\
&=&\ A\{\ g_{\mu\rho}g_{\nu\alpha}g_{\sigma\gamma}g_{\beta\delta}-g_{\mu\rho}g_{\nu\alpha}g_{\sigma\delta}g_{\beta\gamma}-g_{\mu\rho}g_{\nu\beta}g_{\sigma\gamma}g_{\alpha\delta}+g_{\mu\rho}g_{\nu\beta}g_{\sigma\delta}g_{\alpha\gamma}\\
& &\ \ \
-g_{\mu\rho}g_{\nu\gamma}g_{\sigma\alpha}g_{\beta\delta}+g_{\mu\rho}g_{\nu\gamma}g_{\sigma\beta}g_{\alpha\delta}+g_{\mu\rho}g_{\nu\delta}g_{\sigma\alpha}g_{\beta\gamma}-g_{\mu\rho}g_{\nu\delta}g_{\sigma\beta}g_{\alpha\gamma}\\
& &\ \ \ -g_{\mu\sigma}g_{\nu\alpha}g_{\rho\gamma}g_{\beta\delta}+g_{\mu\sigma}g_{\nu\alpha}g_{\rho\delta}g_{\beta\gamma}+g_{\mu\sigma}g_{\nu\beta}g_{\rho\gamma}g_{\alpha\delta}-g_{\mu\sigma}g_{\nu\beta}g_{\rho\delta}g_{\alpha\gamma}\\
& &\ \ \
+g_{\mu\sigma}g_{\nu\gamma}g_{\rho\alpha}g_{\beta\delta}-g_{\mu\sigma}g_{\nu\gamma}g_{\rho\beta}g_{\alpha\delta}-g_{\mu\sigma}g_{\nu\delta}g_{\rho\alpha}g_{\beta\gamma}+g_{\mu\sigma}g_{\nu\delta}g_{\rho\beta}g_{\alpha\gamma}\\
& &\ \ \
-g_{\mu\alpha}g_{\nu\rho}g_{\sigma\gamma}g_{\beta\delta}+g_{\mu\alpha}g_{\nu\rho}g_{\sigma\delta}g_{\beta\gamma}+g_{\mu\alpha}g_{\nu\sigma}g_{\rho\gamma}g_{\beta\delta}-g_{\mu\alpha}g_{\nu\sigma}g_{\rho\delta}g_{\beta\gamma}\\
& &\ \ \
+g_{\mu\beta}g_{\nu\rho}g_{\sigma\gamma}g_{\alpha\delta}-g_{\mu\beta}g_{\nu\rho}g_{\sigma\delta}g_{\alpha\gamma}-g_{\mu\beta}g_{\nu\sigma}g_{\rho\gamma}g_{\alpha\delta}+g_{\mu\beta}g_{\nu\sigma}g_{\rho\delta}g_{\alpha\gamma}\\
& &\ \ \
+g_{\mu\gamma}g_{\nu\rho}g_{\sigma\alpha}g_{\beta\delta}-g_{\mu\gamma}g_{\nu\rho}g_{\sigma\beta}g_{\alpha\delta}-g_{\mu\gamma}g_{\nu\sigma}g_{\rho\alpha}g_{\beta\delta}+g_{\mu\gamma}g_{\nu\sigma}g_{\rho\beta}g_{\alpha\delta}\\
& &\ \ \
-g_{\mu\delta}g_{\nu\rho}g_{\sigma\alpha}g_{\beta\gamma}+g_{\mu\delta}g_{\nu\rho}g_{\sigma\beta}g_{\alpha\gamma}+g_{\mu\delta}g_{\nu\sigma}g_{\rho\alpha}g_{\beta\gamma}-g_{\mu\delta}g_{\nu\sigma}g_{\rho\beta}g_{\alpha\gamma}\}\\
&
&\!\!+B[(g_{\mu\alpha}g_{\nu\beta}-g_{\mu\beta}g_{\nu\alpha})(g_{\rho\gamma}g_{\sigma\delta}-g_{\rho\delta}g_{\sigma\gamma})-(g_{\mu\gamma}g_{\nu\delta}-g_{\mu\delta}g_{\nu\gamma})(g_{\rho\alpha}g_{\sigma\beta}-g_{\rho\beta}g_{\sigma\alpha})]
\end{eqnarray*}
where
\begin{eqnarray*}
A&=&\frac{(D+1)\langle(f^{abc}G^a_{\mu\nu}G^{b}_{\nu\sigma})^2\rangle-\langle(f^{abc}G^a_{\mu\nu}G^b_{\rho\sigma})^2\rangle}{(D+2)D(D-1)(D-2)(D-3)}\\
B&=&\frac{-4\langle(f^{abc}G^a_{\mu\nu}G^{b}_{\nu\sigma})^2\rangle+(D-2)\langle(f^{abc}G^a_{\mu\nu}G^b_{\rho\sigma})^2\rangle}{(D+2)D(D-1)(D-2)(D-3)}\\
\end{eqnarray*}
In 4-dimension:
\begin{displaymath}
f^{abe}f^{cde}\langle0|\widetilde{G}^a_{\mu\nu}(0)\widetilde{G}^b_{\rho\sigma}(0)\widetilde{G}^c_{\alpha\beta}(0)\widetilde{G}^d_{\gamma\delta}(0)|0\rangle=f^{abe}f^{cde}\langle0|G^a_{\mu\nu}(0)G^b_{\rho\sigma}(0)G^c_{\alpha\beta}(0)G^d_{\gamma\delta}(0)|0\rangle\\
\end{displaymath}


\newpage
\begin{thebibliography}{99}


\bibitem{SVZ}
M.A. Shifman, A.I. Vainshtein and V.I. Zakharov, Nucl. Phys. B{\bf
147} 385, 448(1979).
\bibitem{gball1}
V.A. Novikov, M. A. Shifman, A.I. Vainshtein, and Valentin I.
Zakharov, Phys.\ Lett.\ {\bf B86}, 347(1979).
\bibitem{gball2}
V.A. Novikov, M. A. Shifman, A.I. Vainshtein, and Valentin I.
Zakharov, Nucl.\ Phys.\ B{\bf 165},\ 67(1980).
\bibitem{gball3}
M.A. Shifman,\ Z.\ Phys.\ C{\bf 9},\ 347(1981).
\bibitem{gball4}
E. Shuryak, Nucl.\ Phys.\ B{\bf 203},\ 116(1983).
\bibitem{gball5}
S. Narison,\ Z.\ Phys.\ C{\bf 26},\ 209(1984).
\bibitem{gball6}
M.A. Shifman, A.I. Vainshtein and V.I. Zakharov,
Phys.\ Lett.\ {\bf B223}, 251(1989).
\bibitem{gball7}
S. Narison and G. Veneziano, Intern.\ J.\ Mod.\ Phys.\ {\bf A4},\
2751(1989).
\bibitem{gball8}
E. Bagan and T.G. Steele, Phys.\ Lett.\ {\bf B243}, 413(1990).
\bibitem{gball9}
T. Huang, H. J. Jin, and Ailin. Zhang, Phys.\ Rev.\ D{\bf 59},
034026(1999).
\bibitem{gball10}
H. Forkel, Phys.\ Rev.\ D{\bf 64}, 034015(2001).
\bibitem{shuryak}
E. V. Shuryak and J. J. M. Verbaarschot, Phys.\ Rev.\ D{\bf 52},
295(1995).
\bibitem{gball11}
H. Forkel, Phys.\ Rev.\ D{\bf 71}, 054008(2005).
\bibitem{gball12}
S. Narison, Nucl.\ Phys.\ B{\bf 509},\ 312(1998).
\bibitem{gball13}
J. P. Liu, Chin.\ Phys.\ Lett.\ {\bf 15},\ 784(1998).
\bibitem{gball14}
Ailin. Zhang and T. G. Steele, Nucl.\ Phys.\ A{\bf 728}, 165(2003).
\bibitem{lattice1}
D. Weingarten, Nucl.\ Phys. (Proc. Suppl.)\ B{\bf 34},\ 29(1994).
\bibitem{lattice2}
G. Bali, {\it et al.}, Phys.\ Lett.\ {\bf B309},\ 378(1993).
\bibitem{lattice3}
C. Liu, Commun.\ Theor.\ Phys.\ {\bf 35},\ 288(2001).
\bibitem{lattice4}
M. Loan, X. Q. Luo, and Z. H. Luo, {\tt hep-lat/0503038}.
\bibitem{lattice5}
Y. Chen, {\it et al.}, Phys.\ Rev.\ D{\bf 73}, 014516(2006).
\bibitem{narison}
J. Latorre, S. Narison and S. Paban, Phys.\ Lett.\ B{\bf 191},\ 437(1987).
\bibitem{BES}
BES Collaboration, Z. J. Bai, {\it et al}., Phys.\ Rev.\ Lett.\ {\bf
91}, 022001 (2003).
\bibitem{BES2}
BES Collaboration, M. Ablikim, {\it et al}., Phys.\ Rev.\ Lett.\
{\bf 95}, 262001 (2005).
\bibitem{gjding}
G. J. Ding and M.L. Yan, {\tt hep-ph/0511186.}
\bibitem{bali}
B. A. Li, {\tt hep-ph/0510093.}
\bibitem{xghe}
X. G. He, X.Q. Li, and J.P. Ma, {\tt hep-ph/0509140.}
\bibitem{nk}
N. Kochelev and D. P. Min, Phys.\ Lett.\ B{\bf 633},\ 283(2006).
\bibitem{xyshen}
Private communication with Prof. X.Y. Shen.
\bibitem{FDC}
J. X. Wang, Nucl.\ Instrum.\ Methods\ Phys.\ Res.,\ Sect. A{\bf
534}, 241(2004).

\end{thebibliography}
\end{document}